\documentclass[twocolumn,superscriptaddress,aps,preprintnumbers,amsmath,amssymb,prl,nofootinbib]{revtex4-1}

\usepackage{graphicx}
\usepackage{epstopdf}
\usepackage{dcolumn}% Align table columns on decimal point
\usepackage{bm}% bold math
\usepackage{hyperref}
\usepackage{ulem}
\usepackage{color}
\usepackage{bbold}

\begin{document}
%%%%%%%%%%%%%%%%%%%%%%%%%%%%%%%%%%%%%%%%%%%

\def\a{\alpha}
\def\b{\beta}
\def\c{\varepsilon}
\def\d{\delta}
\def\e{\epsilon}
\def\f{\phi}
\def\g{\gamma}
\def\h{\theta}
\def\k{\kappa}
\def\l{\lambda}
\def\m{\mu}
\def\n{\nu}
\def\p{\psi}
\def\q{\partial}
\def\r{\rho}
\def\s{\sigma}
\def\t{\tau}
\def\u{\upsilon}
\def\v{\varphi}
\def\w{\omega}
\def\x{\xi}
\def\y{\eta}
\def\z{\zeta}
\def\D{\Delta}
\def\G{\Gamma}
\def\H{\Theta}
\def\L{\Lambda}
\def\F{\Phi}
\def\P{\Psi}
\def\S{\Sigma}

\def\o{\over}
\def\beq{\begin{eqnarray}}
\def\eeq{\end{eqnarray}}
\newcommand{\gsim}{ \mathop{}_{\textstyle \sim}^{\textstyle >} }
\newcommand{\lsim}{ \mathop{}_{\textstyle \sim}^{\textstyle <} }
\newcommand{\vev}[1]{ \left\langle {#1} \right\rangle }
\newcommand{\bra}[1]{ \langle {#1} | }
\newcommand{\ket}[1]{ | {#1} \rangle }
\newcommand{\EV}{ {\rm eV} }
\newcommand{\KEV}{ {\rm keV} }
\newcommand{\MEV}{ {\rm MeV} }
\newcommand{\GEV}{ {\rm GeV} }
\newcommand{\TEV}{ {\rm TeV} }
\def\diag{\mathop{\rm diag}\nolimits}
\def\Spin{\mathop{\rm Spin}}
\def\SO{\mathop{\rm SO}}
\def\O{\mathop{\rm O}}
\def\SU{\mathop{\rm SU}}
\def\U{\mathop{\rm U}}
\def\Sp{\mathop{\rm Sp}}
\def\SL{\mathop{\rm SL}}
\def\tr{\mathop{\rm tr}}

\def\IJMP{Int.~J.~Mod.~Phys. }
\def\MPL{Mod.~Phys.~Lett. }
\def\NP{Nucl.~Phys. }
\def\PL{Phys.~Lett. }
\def\PR{Phys.~Rev. }
\def\PRL{Phys.~Rev.~Lett. }
\def\PTP{Prog.~Theor.~Phys. }
\def\ZP{Z.~Phys. }

%%%%%%%%%%%%%%%%%%%%%%%%%%%%%%%%%%%%%%%%%%%%%%%%%%%%%%%%%%%%%%%

\title{
Why three generations?
}

\author{Masahiro Ibe}
\affiliation{ICRR, University of Tokyo, Kashiwa, 277-8582, Japan}
\affiliation{Kavli IPMU, University of Tokyo (WPI), Kashiwa, 277-8568, Japan}
\author{Alexander Kusenko}
\affiliation{Kavli IPMU, University of Tokyo (WPI), Kashiwa, 277-8568, Japan}
\affiliation{Department of Physics and Astronomy, University of California, Los Angeles, CA 90095-1547, USA}
\author{Tsutomu T.~Yanagida}
\affiliation{Kavli IPMU, University of Tokyo (WPI), Kashiwa, 277-8568, Japan}

\begin{abstract}
We discuss an anthropic explanation of why there exist three generations of fermions. If one assumes that the right-handed neutrino sector is responsible for both the matter--antimatter asymmetry and the dark matter, then anthropic selection favors three or more families of fermions. 
For successful leptogenesis, at least two right-handed neutrinos are needed,
while the third right-handed neutrino is invoked to play the role of dark matter. The number of the right-handed neutrinos is tied to the number of generations by the anomaly constraints of the $U(1)_{B-L}$ gauge symmetry.
Combining anthropic arguments with observational constraints,
we obtain predictions for the $X$-ray observations, as well as for 
neutrinoless double-beta decay.

\end{abstract}

\date{\today}
\maketitle
\preprint{IPMU16-0014}

\section{Introduction}
One of the most important outstanding questions in modern particle physics is why the Standard Model contains three generations of matter particles.
In the everyday world, it seems that only the first-generation particles (i.e. the electrons and the up and down quarks) play crucial roles.
In view of the simplicity of the fundamental laws of  physics,
the multiple generations of quarks and leptons seem unnecessary.  
This question, succinctly expressed by the famous quip of I.\,I.\,Rabi, {\it ``who ordered that?"}, uttered in connection with the discovery of the muon, has been exacerbated by the discovery of the third generation.    

One possible answer to this question could come from a fundamental theory
which requires three generations as a consistency condition.
For example, such attempts have been made in extra-dimensional
models \cite{Witten:1982fp,Dobrescu:2001ae,Watari:2001qb}. 
The number of generations can also been related to consistency conditions of 
discrete symmetries \cite{Hinchliffe:1992ad,Mohapatra:2007vd,Evans:2011mf}.  However, such extra symmetries or extra dimensions remain  hypothetical. 

In this paper, we will discus an anthropic explanation for the family replication of fermions. As usual, for anthropic selection, we assume that the vacuum of the Standard Model can be realized with a different number of fermion families.   We will show that the three generations are minimal particle content required for existence of life, assuming that both baryon asymmetry and dark matter arise from  the right-handed neutrino sector.  

The existence of right-handed neutrinos is strongly suggested by the measured small neutrino masses, which are naturally explained by the seesaw mechanism~\cite{seesaw}.  Once the right-handed neutrinos with Majorana masses are added to the Standard Model, the matter--antimatter asymmetry can be explained by means of leptogenesis~\cite{Fukugita:1986hr}.  For successful leptogenesis\,\cite{Fukugita:1986hr}, one needs at least two right-handed neutrinos~\cite{Frampton:2002qc}.
 The third one can play the role of dark matter realized as sterile neutrinos~\cite{Dodelson:1993je,Kusenko:2009up}. This fulfils another necessary condition for the existence of life, because structure formation on the relevant length scales requires dark matter. 

In this minimal scenario, in which no new low-energy physics is admitted, the three generations of fermions provide a minimal particle content consistent with the existence of life.  We will also see that cosmological selection not only points to $n_g \ge 3$, but also narrows down the mass range of the sterile neutrinos. 
Furthermore, by imposing existing observational constraints,
we derive predictions for the dark-matter search using X-ray telescopes, as well as for neutrinoless double-beta decay experiments.

\section{Seesaw mechanism} 
Before discussing cosmological selection, let us briefly summarize the seesaw mechanism for $n_g$ generations
of the matter fermion in the Standard Model and $n_N$ Majorana right-handed neutrinos 
\,\cite{seesaw}.
The Lagrangian responsible for the seesaw mechanism is given by,t
\begin{eqnarray}
\label{eq:lagrangian}
 {\cal L} = y_{\a\b}\ell_{L\a} \bar{e}_{R\b}h+ \lambda_{i \a} N_i \ell_{L\a} h^\dagger - \frac{1}{2} M_{Rij} N_i N_j \ .
\end{eqnarray}
Here, $h$ denotes the Higgs doublet,  $\ell_{L\a}\, (\a = 1-n_g)$ the $n_g$ generations of the
lepton doublets,  $\bar{e}_{R\a}\, (\a = 1-n_g) $ the $n_g$ generations of the left-handed anti-leptons,
and $N_i\,(i = 1 -n_N)$ the $n_N$ generations of the right-handed neutrinos.
The coefficients $y$ and  $\lambda$ are the coupling constants and $M$'s are the right-handed neutrino masses.
Hereafter, we take a basis 
where the $M$ and $y$ are diagonal, i.e., $M_{Rij}= M_{Ri} \delta_{ij}$ and $y_{\a\b} = y_{\a}\delta_{\a\b}$.

We assume that the right-handed neutrino mass is generated as a result of 
spontaneous breaking of the $U(1)_{B-L}$ gauge symmetry.
Under this assumption,  the number of the right-handed neutrinos
should be equal to the number of the generations for the anomaly free condition of the $U(1)_{B-L}$
gauge symmetry.
Thus, hereafter, we assume $n_N = n_g$.

By integrating out the heavy right-handed neutrinos, the active $n_g$ neutrino masses are obtained as
\begin{eqnarray}
\label{eq:seesaw}
  (m_\n)_{\a\b}  = \sum_{i=1}^{n_N}\lambda^T_{\a i} M_{Ri}^{-1} \lambda_{i \b}\, v^2\ .
\end{eqnarray}
Here, $v \simeq 174.1$\,GeV is the vacuum expectation value of the Higgs boson.

%%%%%%%%%%%%%%%%%%%%%%%%%%%%%%%%%%%%%%%%%%%%%%%%
\section{Step toward $n_g = 2$ : baryon asymmetry}
To begin with cosmological selection of the number of the generations,
let us briefly review thermal leptogenesis.
For now, let us assume that all of the $n_g$ right-handed neutrinos are heavy
with the lightest right-handed neutrino $N_1$.
When the right-handed neutrino masses are not degenerated, 
the generated baryon asymmetry is proportional to
the $CP$-asymmetry of the decay of $N_1$,
\begin{eqnarray}
 \eta_{B_0} = n_B/n_\gamma \propto \epsilon_1\ .
\end{eqnarray}
 \begin{eqnarray}
 \epsilon_1 = -\frac{3}{16\pi} \frac{M_1}{(\l \l^\dagger)_{11}} 
{\rm Im} [( \lambda \lambda^\dagger M_R^{-1} \lambda^* \lambda^T)_{11}]\ .
\end{eqnarray}
This immediately shows that baryon asymmetry requires $n_g \ge 2$, since $\epsilon_1 = 0$ for $n_g = 1$~\cite{Frampton:2002qc}.%
\footnote{See also \cite{Harigaya:2012bw} for related discussion with a model with $n_g = 2$.}

Without the baryon asymmetry, the relic abundance of the baryons in the universe after freeze-out is highly suppressed.
For such a small baryon density,  disk fragmentation and star formation in a dark halo are, for example, 
precluded \cite{Tegmark:2005dy}.%
\footnote{In this paper, we are not trying to scan all the parameters for cosmological selection
but we only change one of the cosmological parameter (here the baryon asymmetry) and fix 
all the other parameters such as the dark matter density, the  cosmological constant,
and the density perturbation as well as the Standard Model parameters. 
For a more detailed analysis, see Ref.~\cite{Tegmark:2005dy}.
}
The Big-Bang Nucleosynthesis with a scarce baryon density also results in a universe 
with no atoms except  hydrogen which could be detrimental for our existence.
Therefore, for our existence, we find that $n_g \ge 2$ is cosmologically selected 
if we rely on leptogenesis as the origin of baryon asymmetry of the universe.

%%%%%%%%%%%%%%%%%%%%%%%%%%%%%%%%%%%%%%%%%%%%%%%%
\section{Step toward $n_g = 3$ : Dark Matter}
What cosmological selection may lead to $n_g = 3$? 
For $n_g = 3$, we have the third right-handed neutrino which is neutral under the Standard Model gauge group.
Thus, it is quite natural guess that the third right-handed neutrino plays the role of dark matter. 
If this is the case, we can conclude that $n_g = 3$ results from the necessity of dark matter
since dark matter is as crucial as the baryon asymmetry for our existence.%

The observed value of the dark matter to baryon density ratio 
$$
\xi = \frac{\rho_{DM}}{\rho_b} \approx 5.5
$$
is lies in a special range.  Tegmark et al.~\cite{Tegmark:2005dy} show that a combination of several requirements on structure formation and galaxy formation bounds leads to the constraint $2.5\lsim \xi \lsim 10^2 $. The exact bounds depend on the prior assumptions~\cite{Tegmark:2005dy,Hellerman:2005yi}.  At least in the case where one fixes the size of the cosmological constant and the magnitude of the primordial density perturbations, one can show that bounds are  robust. Unless $\xi \gsim 1$, the density perturbations on the galactic scales are washed out by Silk damping before the time of recombination.  These density perturbations can only survive if they are carried by the dark matter, and that requires that the dark matter density dominate over the baryonic density.  If there is too much dark matter, structure forms and goes non-linear very early, before the matter-radiation equality, and the collapsing halos drag the radiation with the baryons, leading to formation of black holes rather than star systems.

Since dark matter is necessary for galaxy formation, anthropic selection strongly favors theories with dark matter candidates.  In our assumed ensemble of theories (or an ensemble of vacua) with different numbers of fermion generations, only $n_g\ge 3$ satisfy the anthropic criteria. In such cases, one of the right-handed neutrinos can be the dark matter. 

Furthermore, the mass of the dark-matter sterile neutrino must be small enough to allow for (i) cosmologically long lifetime, and (ii) acceptable dark matter abundance.   

Interestingly, such a framework has been proposed as ``Split Seesaw" mechanism\,\cite{Kusenko:2010ik},
where the two right-handed neutrinos play curial roles in thermal leptogenesis while the last one plays the role of dark matter.
In this model, the order of the magnitudes of the Yukawa coupling constants $\l$
and the right-handed neutrino masses $M_{R}$ are assumed to be controlled 
by wave function factors of the right-handed neutrinos.
Concretely, $\l$'s and $M_R$'s are given by,
\begin{eqnarray}
\l_{i\a} &\sim& \varepsilon_i\times \tilde{\lambda} \ , \quad (\tilde{\lambda} = {\cal O}(1))\ , \cr
M_{Ri} &\sim& \varepsilon_i^2\times v_{B-L} \ ,
\end{eqnarray}
where $\varepsilon_i$ denotes the suppression factors from the wave functions.
Under this assumption, the light neutrino masses in Eq.\,(\ref{eq:seesaw}) is given by
\begin{eqnarray}
  (m_\n)_{\a\b}  \sim \sum_{i=1}^{3}\tilde{\lambda}^T_{\a i} \tilde{\lambda}_{i \b}\, \frac{v^2}{v_{B-L}}\ ,
\end{eqnarray}
where $\tilde{\lambda}_{i\a }\equiv \lambda_{i\a }/\varepsilon_i$.
The observed neutrino mass splittings, 
${\mit \D}m_{\rm atm}^2 \simeq 2\times 10^{-3}$eV$^2$
and ${\mit \D}m_{\rm sol}^2 \simeq 8\times 10^{-5}$eV$^2$ (see e.g. \cite{Capozzi:2013csa}),
suggest $v_{B-L} \simeq 10^{15}$\,GeV for $\tilde{\lambda} = {\cal O}(1)$ in the split seesaw mechanism.
The closeness of $v_{B-L}$ to the scale of Grand Unified Theory is one of the prime feature of this model.

Now let us discuss whether the third right-handed neutrino is a good candidate for dark matter. 
In the following, we take the mass diagonal base,
\begin{eqnarray}
\nu_s \simeq \nu_{R3} +\sum_{\a=1}^3 \frac{v}{M_{R3}}\lambda_{3\a} \nu_{\a}\ .
\end{eqnarray}

First, let us discuss the lifetime of the sterile neutrino.
The sterile neutrino mainly decays into the three active neutrinos via the above mixing.
The lifetime is given by\,\cite{Pal:1981rm,Gronau:1984rs},
\begin{eqnarray}
\tau_{\nu_s}  \sim 1.4\times 10^{24}{\rm sec} \left(\frac{1\,\rm keV}{M_{R3}}\right)^5\left(\frac{10^{-5}}{\theta_s^2}\right)\ , 
\end{eqnarray}
where the mixing angle is defined by,
\begin{eqnarray}
\theta_s^2 &=& \sum_{\a=1}^3 \left(\frac{v}{m_{s}}\lambda_{3\a} \right)^2 = 
\frac{v^2}{v_{B-L}m_s}\sum_{\a=1}^3\tilde \lambda_{3\a} ^2 \\
&\simeq & 0.3 \times 10^{-4} 
\left(\frac{10^{15}\,{\rm GeV}}{v_{B-L}}\right)
\left(\frac{1\,{\rm keV}}{m_{s}}\right)
\sum_{\a=1}^3\tilde \lambda_{3\a} ^2 
\ .
\end{eqnarray}
As a result, the lifetime of the sterile neutrino can be much longer than the 
age of the universe of ${\cal O}(10^{17})$\,sec, 
with which the sterile neutrino can be a viable candidate for dark matter.%
\footnote{From a view point of anthropic selection,
dark matter with a lifetime shorter than ${\cal O}(10^{17})$\,sec might be acceptable, however, one must still require $\tau_s > {\cal O}(1)$\,Gyr for 
the galaxy formation to take place (assuming the fixed orders of primordial perturbations and cosmological constant.
}

For the sterile neutrino to be an appropriate candidate for dark matter,
the abundance of dark matter should be  in an appropriate range.
In the following, let us discuss the relic abundance of the sterile neutrino.

A population of the sterile neutrinos can be produced through non-resonant 
oscillations  through the mixing to the active neutrino, \cite{Dodelson:1993je}. 
The relic abundance of the sterile neutrino produced by this process is approximated by \cite{Abazajian:2005gj},%
\footnote{Here, we use the QCD transition temperature $T_{\rm QCD}\simeq 170$\,MeV.
See also \cite{Asaka:2006nq} for the estimation of the sterile neutrino abundance.}
\begin{eqnarray}
\Omega_sh^2 \sim 0.12\times
\left( \frac{m_{s}}{3.4\,{\rm keV}}\right)^2
\left(
\frac{\sin^22\theta_3}{10^{-8}}\right)^{1.23}\ .
\label{eq:DW}
\end{eqnarray}
Thus, for example, the observed relic abundance can be explained for $m_s ={\cal O}(1)$\,keV
and $\sin^22\theta_s = {\cal O}(10^{-8})$.

In addition to the production via the non-resonant oscillation, 
the sterile neutrino is also produced from the thermal bath 
of the Standard Model matter fermion via the $U(1)_{B-L}$ gauge interactions.
Since the $U(1)_{B-L}$ gauge interactions are approximated by
dimension six operators suppressed by $v_{B-L}^2$,
this process is most efficient at the reheating epoch after inflation.
The resultant relic abundance is given by~\cite{Kusenko:2010ik},
\begin{eqnarray}
\label{eq:thermal}
\Omega_s h^2 &\sim& 0.13  \times
\left(\frac{m_s}{5\,\rm keV}\right)
\left(\frac{g_*}{100}\right)^{3/2}
\cr
&&
\times
\left(\frac{10^{15}\rm GeV}{v_{B-L}}\right)^{4}
\left(\frac{T_R}{5\times10^{13}\,\rm GeV}\right)^{4}\ .
\end{eqnarray}
Here, $T_R$ denotes the reheating temperature and  
$g_*$ denotes the effective degree of freedom of the massless particles at the reheating temperature.
It should be noted that the reheating temperature which provides the observed dark matter density
is $T_R = {\cal O}(10^{13})$\,GeV, which is also appropriate for successful leptogenesis, 
$T_R \gtrsim 10^{10}$\,GeV\,\cite{Fukugita:1986hr}.

Finally, the sterile neutrino is also produced via the weak interaction through the mixing to the active neutrinos
which is most efficient at the electroweak symmetry transition.
The relic abundance from this process is roughly given by
 \begin{eqnarray}
\Omega_s h^2 &\sim& 0.1  \times
\left(\frac{m_s}{5\,\rm keV}\right)\left(\frac{\sin^22\theta_s}{10^{-6}}\right)\ ,
\end{eqnarray}
for the sterile neutrino much lighter than the weak scale. 
The relic abundance from this process is subdominant compared with the one from the non-resonant oscillation.  The production of sterile neutrinos at the electroweak scale can be enhanced if they couple to a Higgs singlet~\cite{Kusenko:2006rh,Petraki:2007gq}.  The addition of such a singlet would represent extra structure on top of the minimal split seesaw model we consider. 

The sterile neutrino with a mass in the keV range can have a non-negligible free-streaming length (whose actual value depends on the production scenario~\cite{Kusenko:2009up,Kusenko:2006rh,Asaka:2006ek,Patwardhan:2015kga}). which affects cosmological selection
of the sterile neutrino dark matter because structure formation is suppressed on scales smaller than the free-streaming length~\cite{Bond:1980ha}.
For dark matter produced by non-resonant neutrino oscillations, this corresponds to the comoving length\footnote{Here, we define the comoving  free-streaming length by $\l_{FS} = 2\pi/k_{FS}$, where
\begin{eqnarray}
k_{FS} = \left(\frac{3 a^2H_0^2 \Omega_{\rm DM}}{2 \vev{v^2}}  \right)^{1/2}\ .
\end{eqnarray}
}
\begin{eqnarray}
\lambda_{FS} \sim 0.8 \,{\rm Mpc} \left(\frac{1\,\rm keV}{m_s}\right)\ .
\end{eqnarray}

If the structures smaller than the tens to the hundreds Mpc are erased, no mass structures
go nonlinear in the universe, which precludes the galaxy/star formation.
Thus, cosmological selection puts a lower bound on the sterile neutrino 
mass around ${\cal O}(10^{-(1-2)})$\,keV.

In addition, the sterile neutrino is fermionic dark matter, and hence, its phase space density is limited from above.
For the sterile neutrino whose distribution is given by the Fermi-Dirac distribution, 
the upper limit on the phase space density leads to a lower limit on the sterile neutrino mass \cite{Gorbunov:2008ka},
\begin{eqnarray}
m_s \gtrsim 5\,{\rm keV} \times\left(\frac{q}{5\times 10^{-3}}\right)^{1/3}\ .
\end{eqnarray}
Here, $q$ denotes the phase space density estimated from 
the ratio between the mass density and cube of the velocity dispersion $\sigma$\,\cite{Gorbunov:2008ka}
\begin{eqnarray}
Q = \frac{\rho }{\sigma^3} = q \frac{M_{\odot}/pc^3}{({\rm km}/{\rm sec})^3} \ .
\end{eqnarray}
Thus, for example, if we impose cosmological selection so that the universe has at least 
the non-linear structure of the size of the galaxy cluster ($q = {\cal O}(10^{-13}$), 
the sterile neutrino mass should be larger than ${\cal O}(1)$\,eV.
This constraint is weaker than the one from the free-streaming length.

%%%%%%%%%%%%%%%%%%%%%
\begin{figure}[t]
\begin{center}
  \includegraphics[width=.8\linewidth]{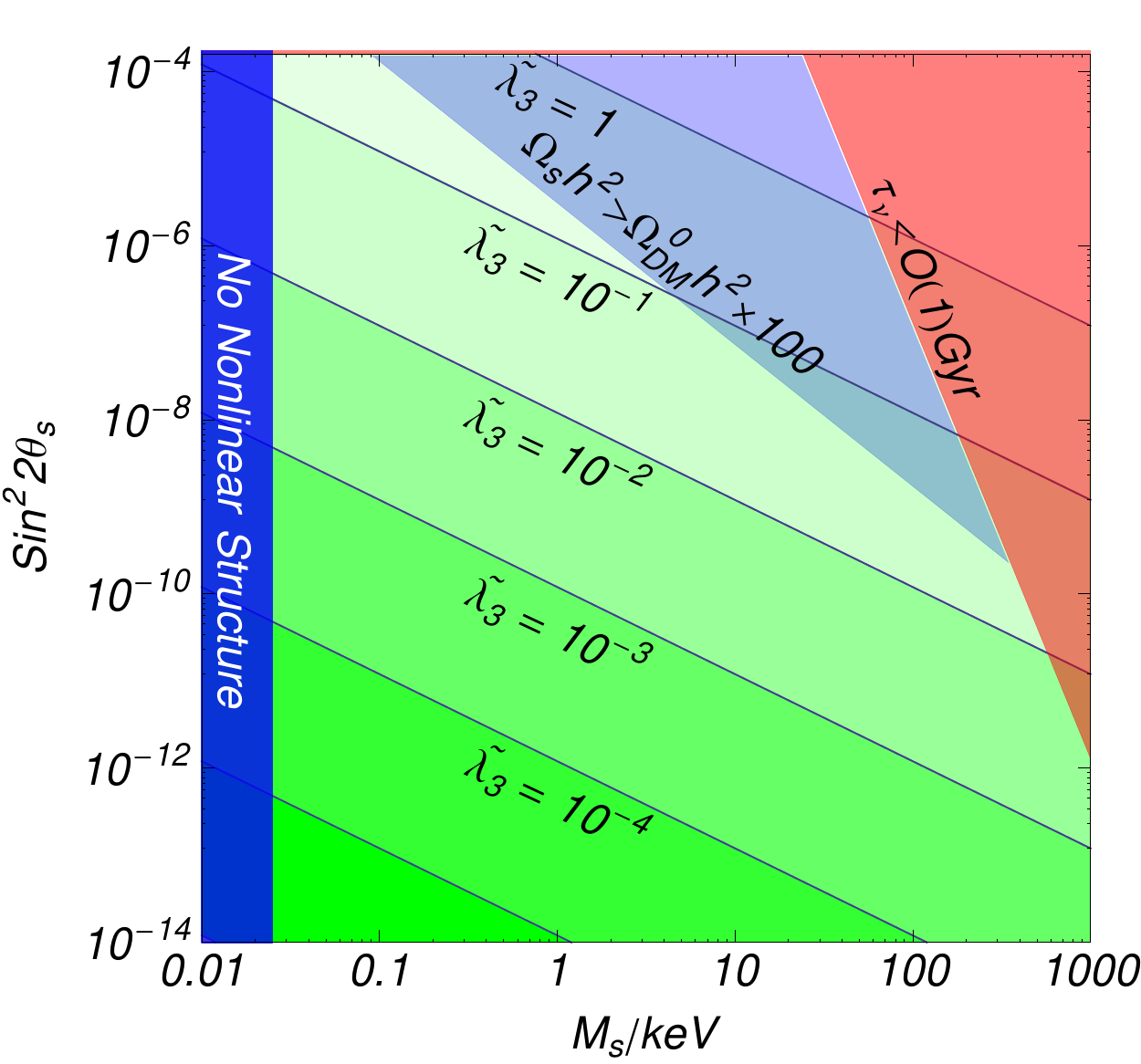}
 \end{center}
\caption{\sl \small
The parameter region of the sterile neutrino dark matter 
which survives the cosmological selection.
The red shaded region is excluded where the lifetime of 
dark matter is shorter than ${\cal O}(1)$\,Gyr which precludes the first
galaxy formation.
The blue shaded region is excluded where no mass structure in the universe 
goes into nonlinear, $m_s >{\cal O}(10^{-(1-2)})$\,keV.
The light-blue shaded region corresponds to the too high dark matter density, 
$\Omega_sh^2 > \Omega_{\rm DM}^0 h^2 \times 100$.
We also show the contour plots of the corresponding value of $\tilde\lambda_3$.
}
\label{fig:CS}
\end{figure}
%%%%%%%%%%%%%%%%%%%%%

Now, let us summarize the parameter region of the sterile neutrino dark matter 
which survives the cosmological selection.
In Figure\,\ref{fig:CS}, we show the surviving region on the $(m_s, \sin^22\theta_s)$ plane.
The red shaded region is excluded 
where the lifetime of 
dark matter is shorter than ${\cal O}(1)$\,Gyr which precludes the first
galaxy formation.
According to \cite{Tegmark:2005dy}, we also exclude the region where 
the dark matter density (in Eq.\,(\ref{eq:DW})) is larger than the observed density, $\Omega_{\rm DM}^0 h^2\simeq 0.12$ 
by a factor of a hundred since such a dense dark matter precludes the disk formation of the Milky way type galaxies.
In other region, we require appropriate reheating temperature to provide an appropriate dark matter density (see
Eq.\,(\ref{eq:thermal})).
The region with $m_s < 10^{-(1-2)}$\,keV is  also excluded where no mass structure in the universe 
goes into nonlinear.

We conclude that the sterile neutrino dark matter, which appears for $n_g = 3$, survives the cosmological selection.
This, in turn, shows that the three generations can be a result of cosmological selection for the necessary amount of dark matter.
Furthermore, the above argument of the cosmological selection 
does not only lead to $n_g = 3$ but also narrows the range of the sterile neutrino mass range.
Besides, if one tries to avoid a severe fine-tuning in the split seesaw mechanism, the Yukawa coupling $\lambda_3$ should not be very small, (e.g. $\lambda_3 > {\cal O}(10^{-(2-3)})$),
which predicts the mass of the sterile neutrino in $m_s \sim 10^{-2}$\,keV--$10^{2}$\,keV.%
\footnote{For a model which provides a small $\tilde\lambda_3$, see \cite{Ishida:2013mva}. }

\section{Observational Constraints and Predictions}
\begin{figure}[t]
\begin{center}
  \includegraphics[width=.8\linewidth]{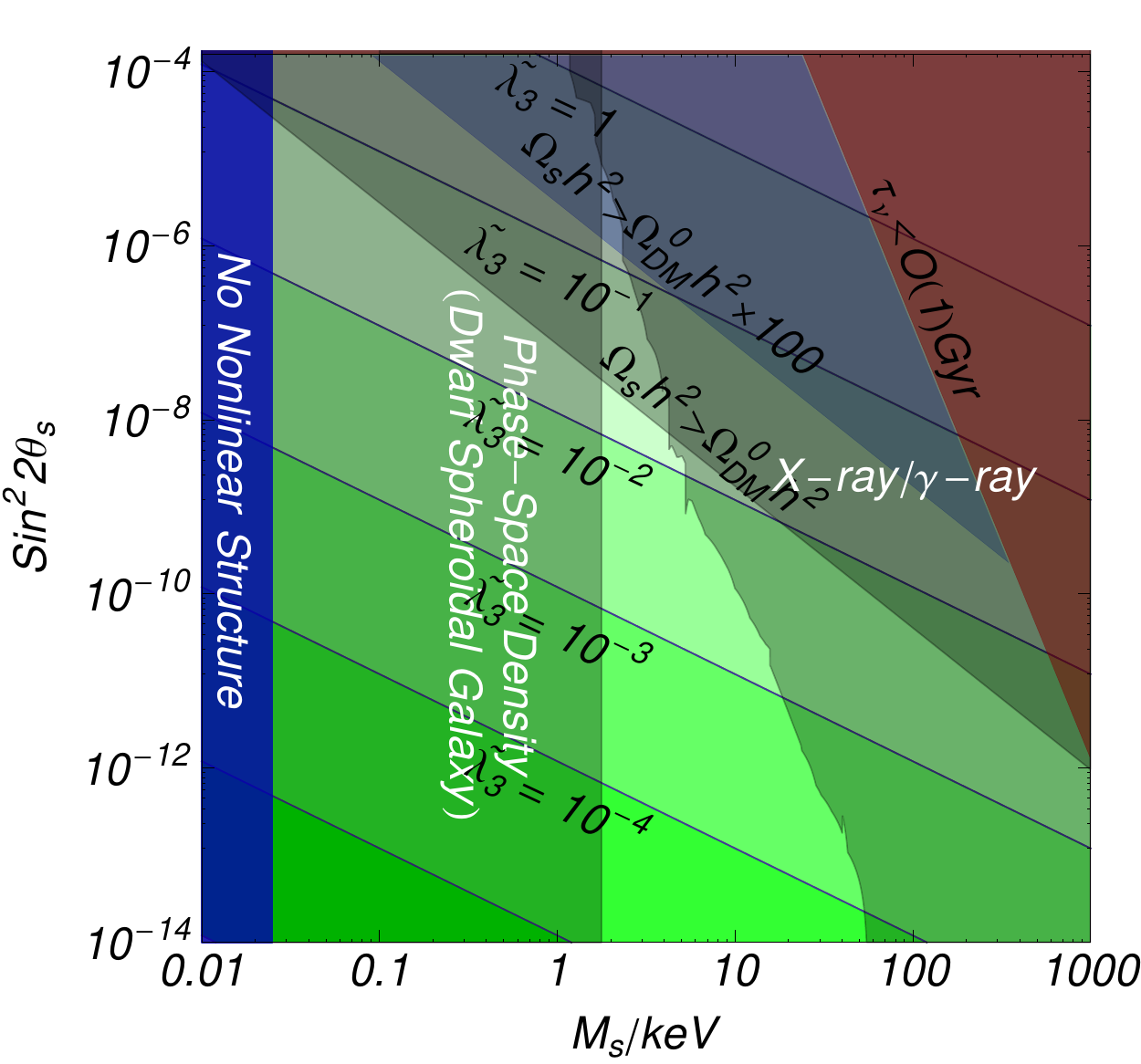}
 \end{center}
\caption{\sl \small
The same figure of Fig.\,\ref{fig:CS} with the observational constraints overlayed.
}
\label{fig:OB}
\end{figure}

Let us juxtapose the observational constraints on the sterile neutrino dark matter with those which resulted from anthropic consideration (Fig.~\ref{fig:OB}).
First of all, the region where the dark matter density exceeds the observed density is excluded.
Various $X$-ray/$\gamma$-ray observations also put constraints on the mass and the mixing angle
of the sterile 
neutrinos\,\cite{Boyarsky:2005us,Boyarsky:2006zi,Watson:2006qb,Boyarsky:2006ag,
Boyarsky:2007ge,Loewenstein:2008yi,Loewenstein:2009cm,Loewenstein:2012px,Ng:2015gfa}. 
The phase-space considerations of the dwarf spheroidal galaxies in the Milky Way also exclude the sterile neutrino mass below $5.7$\,keV\,\cite{Gorbunov:2008ka} if all dark matter is produced from non-resonant neutrino oscillations.%
\footnote{In the parameter space of our interest, 
the sterile neutrinos are dominantly produced at high temperature regions as given in Eq.\,(\ref{eq:thermal}),
and decouple from thermal bath immediately.
Thus, the dark matter momenta are red-shifted because of the entropy production in the Standard Model
which leads to a shorter free-streaming length by a factor about $3$.
In Fig.\,\ref{fig:OB}, we show the lower limit on the mass divided by this factor.}  
The constraint from Lyman-$\alpha$ forest also exclude the sterile mass below $2.5$\,keV\,\cite{Seljak:2006qw}.%
The constraints from the phase-space density and the small scale structure are  weakened or eliminated in Split Seesaw model~\cite{Kusenko:2010ik} or other models in which dark matter is produced at temperatures above the QCD transition~\cite{Kusenko:2006rh,Petraki:2007gq,Petraki:2008ef}, because the entropy production leads to red-shifting of dark matter velocities.  The dark matter can also be cooled by entropy production from decays of additional particles~\cite{Patwardhan:2015kga}.
Those excluded regions are shaded by gray. 

Altogether, we find that the large portion of the survived parameter region has been excluded by
the observational constraints.
In addition, by disfavoring a severely fine-tuned parameter region in the split seesaw mechanism, $\tilde \lambda_3 \ll {\cal O}(10^{-2})$,
we obtain a sharp prediction on the sterile neutrino mass from around $2$\,keV to a few $10$\,keV.

Interestingly, this region includes $m_s \simeq 7.1$\,keV and $\sin^22\theta_s \simeq 7\times 10^{-11}$
which can explain the $X$-ray line signals at a photon energy of around $3.55$\,keV from
various sources\,\cite{Bulbul:2014sua,Boyarsky:2014jta,Boyarsky:2014ska}.
At this point, the existence of this signal is still under debate (see e.g. \cite{Urban:2014yda}),
which can be settled by future $X$-ray telescopes such as ASTRO-H and ATHENA.

As another interesting prediction, the allowed region corresponds to 
slightly fine-tuned Yukawa coupling, i.e. $ \tilde\lambda_3 \sim 10^{-2} $.
Therefore, the contribution of the third right-handed neutrino to the 
active neutrino mass is highly suppressed, which 
leads to the lightest active neutrino mass of ${\cal O}(10^{-6})$\,eV.%
\footnote{Here, we fix $v_{B-L}\simeq 10^{15}$\,GeV and $\tilde \lambda_{1,2} = {\cal O}(1)$.
If we allow $\tilde \lambda_{1,2} \ll {\cal O}(1)$ and $v_{B-L}\ll 10^{15}$\,GeV,
the third right-handed neutrino may have sizable contributions to the active neutrino masses.
}
For such a small lighter active neutrino mass, 
the effective Majorana neutrino mass for the neutrinoless double beta decay
is predicted to be $m_{ee} \simeq 1$\,meV--$5$\,meV for the normal hierarchy
and $m_{ee} \simeq 20$\,meV--$50$\,meV for the inverted hierarchy (see e.g. \cite{Benato:2015via}).
The sensitivity of current experiments is at the $100$\,meV level 
by using with fiducial $\b\b$ masses of around $100$\,kg of Xe
\cite{Albert:2014awa,Asakura:2014lma}.
Thus, for the inverted hierarchy, it is possible to test the model by upcoming 
experiments such as upgraded KamLand-Zen, SNO+, CANDLES, and AMoRE.
For the normal hierarchy, on the other hand, the predicted effective Majorana neutrino mass 
requires about hundred times larger detectors are required to be tested.

%%%%%%%%%%%%%%%%%%%%%%%%%%%%%%%%%%%%
%%%%%%%%%%%%%%%%%%%%%%%%%%%%%%%%%%%%

\section{Conclusion}
We have shown that anthropic selection favors three or more generations of fermions if one assumes that 
the right-handed neutrino sector provides the baryon asymmetry,
via leptogenesis, and also the dark matter.  

For successful leptogenesis, one needs at least two right-handed neutrinos,
while the third right-handed neutrino plays the role of (sterlie neutrino) dark matter.  The number of the right-handed neutrinos is tied to the number 
of generations of fermions in the Standard Model via
the anomaly cancellation condition of the $U(1)_{B-L}$ gauge symmetry.

We also found that the mass of the sterile neutrino dark matter 
is predicted to be in the range of $m_s \sim 10^{-2}$\,keV--$10^{2}$\,keV
by cosmological selection and a requirement to minimize fine-tuning.
The existing constraints exclude a large portion of the parameter space which survived the cosmological selection, rendering a sharp prediction on the mass
and the mixing angle of the sterile neutrino.
Our scenario can be tested by future $X$-ray observations and the searches for
the neutrinoless double beta decay.

\section{Acknowledgements}
This work is supported in part by Grants-in-Aid for Scientific Research from the Ministry of Education, Culture, Sports, Science, and Technology (MEXT) KAKENHI, Japan, No. 24740151, No. 25105011 
and No. 15H05889 (M.~I.) as well as No. 26104009 (T.~T.~Y.); Grant-in-Aid No. 26287039 (M.~I. and T.~T.~Y.) from the Japan Society for the Promotion of Science (JSPS) KAKENHI; and by the World Premier International Research Center Initiative (WPI), MEXT, Japan (M.~I., and T.~T.~Y.).
This work is also supported by MEXT Grant-in-Aid for Scientific research on Innovative Areas (No.15H05889).
The work of A.K. was supported by the U.\,S.\ Department of Energy Grant DE-SC0009937.

\renewcommand{\bibsection} {
\end{document}